\documentstyle[preprint,eqsecnum,tighten,prd,aps]{revtex}
\newcommand{\beq}{\begin{equation}}
\newcommand{\eeq}{\end{equation}}
\newcommand{\bea}{\begin{eqnarray}}
\newcommand{\eea}{\end{eqnarray}}
\def\t{\tilde}
\def\tN{\tilde N}
\def\g{\gamma}

\def\tP{\tilde \Pi}
\begin{document}
\draft
\preprint{\vtop{\hbox{APCTP 97-14}\hbox{hep-th/9708093}}}
\title{Hamiltonian formulation of the D-brane action \\ and the light-cone 
Hamiltonian}
\author{Julian Lee\thanks{\noindent e-mail: jllee@dirac.hanyang.ac.kr}}
\address{ Research Institute for Natural Sciences\\ Hanyang University\\ Seoul 
133-791, Korea}
\date{October 27, 1997}
\maketitle
\begin{abstract}
We present the Hamiltonian formulation of the bosonic Dirichlet $p$-brane 
action. We rewrite the recently proposed quadratic D-brane action in terms of 
generalized shift vector and lapse function. The first class and the second 
class constraints are explicitly separated for the bosonic case. We then impose  
the gauge conditions in such a way that only time-independent gauge 
transformations are left. In this gauge we obtain the light-cone Hamiltonian 
which is quadratic  in the field momenta of scalar and vector fields. The 
constraints are explicitly solved to eliminate part of the canonical variables. 
The Dirac brackets between the remaining variables are computed and shown to be 
equal to simple Poisson brackets.  
\end{abstract}

\section{Introduction}
The action for D-brane which is quadratic in derivatives of $X$ and linear in 
$F_{ij}$ was introduced in ref.\cite{Lin} and recently discussed in 
ref.\cite{ZH}, which takes the form
\begin{equation}
S = - \frac{1}{2}T_p \int d^n \xi \sqrt{-k} \left[ ( k^{-1})^{i j } \left( 
\partial_i X^\mu \partial_j X_\mu +
F_{i j} \right)-(n-2) \right] .
\label{EH}
\end{equation} 
where we are considering a bosonic D-$p$-brane propagating in a $D$ dimensional 
flat background for simplicity\footnote{The notation for the target space 
dimension, $D$, has nothing to do with D in D-$p$-brane, which is just an 
abbreviation of Dirichlet.}, with $\mu=0, \cdots D-1$, and $T_p$ is the 
$p$-brane tension, with $p=n-1$. Here the auxiliary field $k_{ij}$ is introduced 
so as to remove the square root from the Born-Infeld type action for 
D-branes\cite{CS,AS,Doug,PKT,Ark,Ceral,JSetal,GHT,BT},
\begin{equation}
S_{DBI} = -T_p \int d^n \xi \sqrt{-\det{\left( \partial_i X^\mu \partial_j X_\mu 
+F_{i j} \right) }} ,
\label{GDBI}
\end{equation} 
Thus the action given by eq.(\ref{EH}) is the analogue of the Polyakov action in 
the case of string\cite{Polya,BVH} (or corresponding generalization for higher 
dimensional branes\cite{HTU}), which is the linearization of the Nambu-Goto 
action. The only difference from ordinary $p$-brane is that due to the presence 
of the gauge field strength $F_{i j}$, the auxiliary field $k_{ij}$ must also 
contain antisymmetric part.
   
   Gauge-fixing of D-brane using the  static gauge was discussed in 
ref.~\cite{JSetal}, where they started from the Born-Infeld type of action 
(\ref{GDBI}) and arrived at a gauge-fixed action which is a complicated 
non-linear action. Also a covariant formalism  was studied in 
ref.~\cite{Ka}. Derivation of the first-order action (\ref{EH}) from the 
Born-Infeld type action (\ref{GDBI}) using a Hamiltonian description was also 
discussed in ref.~\cite{LU}. Classical symmetries of the action (\ref{EH}) for 
$n=2$ (D-string) was investigated in ref.~\cite{BS}.
 
 In this paper, I concentrate my attention to the canonical Hamiltonian 
formalism. At the classical level, the Hamiltonian analysis has an advantage of 
exhibiting the relevant physical degrees of freedom more clearly. When 
one attempts to go to quantum theory, it can be directly connected with the 
operator formalism via Dirac quantization. Since the action given by 
(\ref{GDBI}) and (\ref{EH}) are classically equivalent, one can apply this 
formalism to either one, but I chose to use the linear form (\ref{EH})  since it 
is more convenient for manipulations such as separating primarily inexpressible 
velocities and fixing gauges, and so on. It turns out that when one uses the 
Born-Infeld type action (\ref{GDBI}), some complicated expressions appear due to 
the presence of square root. Similar situation arises in case of string or 
membrane.  We then go to the light-cone gauge to obtain the light-cone 
Hamiltonian which is quadratic in field momenta. This is the generalization of 
the light-cone quantization of ordinary string\cite{GS} and membrane\cite{BST}.    

\section{The constraints in the Hamiltonian formalism}
 To go to the Hamiltonian formalism, it is convenient to parameterize the 
auxiliary field $k_{i j}$ in terms of generalized shift vectors $N_a$, 
$\tN_a$ and lapse function $N$ as follows,
\bea
k_{00}&=&-N^2+\gamma_{a b} N^a \tN^b, \> k_{0a}=N^b \gamma_{b a} \nonumber \\
k_{a0}&=& \gamma_{ab} \tN^b,\quad  k_{a b}=\gamma_{a b} \nonumber \\
k^{00} &=& -N^{-2},\quad  k^{0a}=N^a N^{-2} \label{sl} \\
k^{a0} &=& \tN^a N^{-2}, \quad k^{ab} = \gamma^{ab} - \tN^a N^b N^{-2} 
\nonumber \\
\sqrt{-k}&=& N\sqrt{\gamma} \nonumber
\eea  
where $\gamma_{a b} \ (a,b=1,\cdots,n-1)$ are the spatial components of the 
auxiliary field and $\gamma^{a b}$ is the inverse, defined by  $\gamma^{a 
b}\gamma_{b c} = \delta^a_c$. In terms of these variables, the action 
(\ref{EH}) is written as,
\bea
S &=&\! T_p \int d^n \xi  [  {\sqrt{ \gamma} \over 2N} {\dot X}^2 - 
{\sqrt{\gamma} \over 2N}(N^a + \tN^a) \dot X \cdot \partial_a X - 
{\sqrt{\gamma}\over 2N} (N^a-\tN^a) F_{0a} \nonumber \\ &&-{\sqrt{\g} \over 
2}(\gamma^{a b} N - { \tN^a N^b \over N})(\partial_a X \cdot \partial_b X 
\!+\!F_{ab} ) + {(n-2)\over 2}  N\sqrt{\gamma} ] \nonumber \\
&\equiv& T_p \int d^n \xi {\cal L}
\eea  
 The canonical variables are $(X^\mu,N, N^a, \tN^a, \g^{ab}, A_i)$ and 
their conjugate momenta $(P_\mu, \Pi, \Pi_a, \tP_a, \Pi_{a b}, \pi^i)$, which 
are given by 
\begin{mathletters}
\label{canmom}
\bea
P_\mu &\equiv& {\partial {\cal L} \over \partial \dot X^\mu } = {\sqrt{\gamma} 
\over N 
} \dot X_\mu - {\sqrt{\gamma} \over 2 N}(N^a + \tN^a) \partial_a X_\mu  
\label{mom1} \\
\Pi &\equiv& {\partial {\cal L} \over \partial \dot N} = 0 \label{mom2}  \\
\Pi_a &\equiv& {\partial {\cal L} \over \partial \dot N^a} = 0   \\   \tilde 
\Pi_a 
&\equiv& {\partial {\cal L} \over \partial \dot {\tilde N^a}} = 0  \\
\Pi_{a b} &\equiv& {\partial {\cal L} \over \partial \dot \g^{a b}} = 0   \\ 
\pi^0 &\equiv&  {\partial {\cal L} \over \partial \dot A_0} = 0 \label{mom6} \\
\pi^a &\equiv&  {\partial {\cal L} \over \partial \dot A_a} = -{\sqrt{\g}\over 
2N}(N^a - \tN^a) \label{mom7}  
\eea   
\end{mathletters}
The velocity $\dot X^\mu$ is a {\it primary expressible velocity}, meaning that 
 it can be solved in terms of momenta, whereas the other ones are 
primary inexpressible and give rise to the {\it primary 
constraints}\cite{GT,HT,D}
\begin{mathletters}
\label{prcon}
\bea
\Omega &\equiv& \Pi \approx 0 \label{p1}\\
\Omega_a &\equiv& \Pi_a \approx 0 \\
\tilde\Omega_a &\equiv& \tilde\Pi_a \approx 0 \\
\Omega_{a b} &\equiv& \Pi_{a b} \approx 0 \\
\phi^0 &\equiv& \pi^0 \approx 0\\
\phi^a &\equiv& \pi^a + {\sqrt{\g} \over 2N}(N^a-\tilde N^a) \approx 0 
\label{p6}
\eea 
\end{mathletters}
 The notation $\approx 0$ indicates that these constraints are weakly 
zero, meaning that they are numerically restricted to be zero but do not 
identically  vanish throughout phase space. This means, in particular, that they 
have nonzero 
Poisson brackets with the canonical variables\cite{GT,HT,D}. The time evolution 
of the system is generated by the Hamiltonian
\begin{mathletters}
\bea
H^{(1)} &=& \int d^{n-1} \xi [ P_\mu \dot X^\mu -L + \Sigma \Omega + \Sigma^a 
\Omega_a + \tilde \Sigma^a \tilde \Omega_a + \Sigma^{a b} \Omega_{a b} + 
\lambda^0 \phi_0 + \lambda^a \phi_a] \nonumber \\
&=& \int d^{n-1} \xi [ {N \over 2\sqrt{\g}}P_\mu P^\mu + {N \sqrt{\g} \over 2} 
(\g^{a b}+ {n_-^a n_-^b \over 4N^2}) \partial_a X^\mu \partial_b X_\mu  
\nonumber \\
& & \quad - {N \sqrt{\g} \over 2}(n-2)+{n_+^a \over 2} \partial_a X^\mu P_\mu - 
 {\sqrt{\g}\over 2N}n_-^a \partial_a A_0 \nonumber \\
& & \quad + {\sqrt{\g}\over 2} (N\g^{a b} - {\tN^a N^b \over N}) F_{a b} + 
\Sigma \Omega + \Sigma^a \Omega_a + \tilde \Sigma^a \tilde \Omega_a + \Sigma^{a 
 b} \Omega_{a b} \nonumber \\
& & \quad + \lambda^0 \phi_0 + \lambda^a \phi_a] \\ 
&\equiv& H + \int[\Sigma \Omega + \Sigma^a \Omega_a + \tilde \Sigma^a \tilde 
\Omega_a + \Sigma^{a b} \Omega_{a b}+ \lambda^0 \phi_0 + \lambda^a \phi_a] 
\label{ham}
\eea  
\end{mathletters}
where $n_\pm^a \equiv N^a \pm \tN^a$, i.e. for any canonical variable $\eta$, we 
have
\beq
\dot \eta = \{ \eta, H^{(1)} \} \label{ph}
\eeq
on the constraint surface, where  the Poisson bracket of any two 
functions $F(\{\eta\})$ and $G(\{\eta\})$ is defined by
\bea
\{ F, G \} &\equiv& \int d^p \xi [ {\delta F \over \delta X^\mu}{\delta G \over 
 \delta P_\mu}+ {\delta F \over \delta N}{\delta G \over \delta \Pi}+ {\delta 
F \over \delta N^a}{\delta G \over \delta \Pi_a}+ {\delta F \over \delta 
\tN^a}{\delta G \over \delta \tP_a}+ {\delta F \over \delta \g^{a b}}{\delta 
G \over \delta \Pi_{a b}} \nonumber \\
& &+ {\delta F \over \delta A^0}{\delta G \over \delta \pi_0}+ {\delta F \over 
\delta A^a}{\delta G \over \delta \pi_a}- ( F \leftrightarrow G ) ]. 
\eea 
 Here, $\Sigma, \Sigma^a, \tilde \Sigma^a, \Sigma^{a b}, \lambda^0, \lambda^a$ 
 are the Lagrange multipliers and they represent the primary 
inexpressible velocities. We must now require the primary constraints to be 
maintained in time,
\beq
\{ \Phi^{(1)}, H^{(1)} \} \approx 0 \label{evo}.
\eeq 
 The relation (\ref{evo}) determines the Lagrange 
multipliers $\Sigma^a-\tilde\Sigma^a, \lambda_a$, and additionally they give 
rise to second stage constraints,
\begin{mathletters}
\label{seccon}
\bea
\varphi &\equiv& {1 \over 2} P^2 + {\g \over 2} (\g^{c d} + {n_-^c n_-^d \over 
4N^2} ) \partial_c X^\mu \partial_d X_\mu -{\g \over 2}(n-2) \nonumber \\
& &+ {\g \over 2} \g^{c d} F_{c d} \label{s1}\\
\varphi_a &\equiv& P_\mu \partial_a X^\mu -{\sqrt{\g} \over 2N}(N^b- \tN^b) 
F_{ab} \\
\varphi_{a b} &\equiv& \partial_a X^\mu \partial_b X_\mu + F_{b a}-\g_{b a} 
\\
\bar \phi &\equiv& \partial_a[{\sqrt{\g} \over 2N}(N^a - \tN^a)] \label{s4}
\eea  
\end{mathletters}
By requiring these constraints to be maintained in time, i.e. their 
Poisson brackets with $H^{(1)}$ vanish weakly, one determines the Lagrange 
multiplier $\Sigma_{a b}$, but no new constraint appears. Therefore, there is no 
third stage constraint and the second stage constraints above are all the 
secondary constraints we have. Note that the Lagrange multipliers $\Sigma, 
\Sigma_a+\tilde \Sigma_a, \lambda_0$ are still undetermined. Since these are 
primary inexpressible velocities, this means that there are arbitrary functions 
of time  and the time evolution of the system is not unique. Of course, this is 
a consequence of the gauge symmetry of the system, and any arbitrariness 
is associated with the unphysical degrees of freedom.

\section{The first and second-class constraints}
 Now, the constraints can be divided into first-class and 
second-class constraints. A first-class constraint $\Psi_f$ is the one whose 
Poisson bracket  with any other constraint $\Phi$ vanishes weakly, $\{\Phi_f, 
\Phi\} \approx 0$. On the other hand, the second-class constraints $\chi_l$ are 
the ones whose Poisson brackets among them form a nonsingular matrix, $\det 
\{\chi_l, \chi_{l'}\} \not \approx 0$. For bosonic variables, this matrix is 
antisymmetric, and therefore the number of second-class constraints should be  
even~\cite{GT,HT}. After some long and tedious calculations, we find 
them to be given by:

\begin{mathletters}
\label{fcon} 
\noindent {\it First-class constraints} 
\bea
\bar \Omega &\equiv&  \Pi + {(N^a-\tN^a) \over 2N}(\Pi^a-\tP^a)  \label{f1} \\
\Omega^+_a &\equiv& \Pi_a + \tP_a  \label{f2} \\
\phi^0 &\equiv& \pi^0  \label{f3} \\
\phi &\equiv& \partial_a  \phi^a - \bar \phi = \partial_a \pi^a   \label{f4} \\ 
 T &\equiv& \varphi + 2 \Omega^{c d} P \cdot \partial_c \partial_d X + P \cdot 
\partial_c X \partial_d (\Pi^{c d} + \Pi^{d c}) \nonumber \\
& &- {\sqrt{\g} \over 2N} \partial_l X \cdot \partial_m X n_-^m \phi^l 
+[{\g^{bc} \over 2}P\cdot \partial_b \partial_c X n_-^d \nonumber \\
&& + {N P \cdot \partial_b X \over \sqrt{\gamma}}\partial_c\{ 
{\sqrt{\g}(\g^{bc}+\g^{cb})n_-^d \over 4N} \} \nonumber \\
&&- {n_-^l \over 8} \partial_l X\cdot \partial_m X \partial_c\{ 
{\sqrt{\g}(\g^{cm}+\g^{mc}) n_-^d \over N} \}](\Pi_d - \tP_d) \nonumber \\
&& + [ {\g \over 2}(\g^{cd}-\g^{dc}) + {\sqrt{\g} \over 4 N} (\g^{bc}+\g^{cb}) 
n_-^d P \cdot \partial_b X  \nonumber \\
&&- {\g n_-^l \over 8N^2} \partial_lX\cdot \partial_m X   
(\g^{cm}-\g^{mc})n_-^d]\partial_c[{N \over \sqrt{\g}}(\Pi_d-\tP_d)] \label{f5} 
\\
T_a &\equiv& \varphi_a + \partial_c[(\Pi^{cd}+\Pi^{dc})\partial_d X ]\cdot 
\partial_a X -{\sqrt{\g} \over 2N} n_-^c\partial_a[{N \over \sqrt{\g}}(\Pi_c - 
\tP_c)] \nonumber \\
 &&+{\sqrt{\g} \over 2N} n_-^c \partial_c[{N \over \sqrt{\g}}(\Pi_a - \tP_a)] 
+\partial_c[ {(\g^{cb} + \g^{bc})n_-^d \over 4}\partial_b X (\Pi_d - \tP_d) ] 
\cdot \partial_a X \nonumber \\
 &&+F_{al}\partial_c[ {(\g^{cl}-\g^{lc})n_-^d \over 4 }(\Pi_d -\tP_d)] + F_{al} 
 \phi^l \nonumber \\
&&+ F_{al} \partial_k (\Pi^{lk}-\Pi^{kl}) \label{f6} 
\eea 
\end{mathletters} 
\noindent {\it Second-class constraints}
\begin{mathletters}
\label{secon}
\bea
\Omega_{ab} &\equiv& \Pi_{a b}  \label{dnt} \\
\varphi_{a b} &\equiv& \partial_a X^\mu \partial_b X_\mu + F_{b a} - \gamma_{a 
b} \quad  (n \neq 2) \qquad \qquad \qquad \qquad \label{dntt}\\ 
\Omega^-_a &\equiv& \Pi_a - \tP_a  \label{dnth} \\
\phi^a &\equiv& \pi^a + {\sqrt{\g} \over 2N} (N^a - \tN^a)  \label{yes}
\eea  
\end{mathletters} 
The constraint (\ref{dntt}) appears only for $n\neq 2$. In this case, 
the constraint (\ref{dnt}) is a first-class constraint. Once any set of 
second-class constraints $\{\chi_l\}$ (which do not explicitly depend on time) 
are found, the time evolution of the system is given by 
\beq
\dot \eta = \{ \eta, H + \tilde \lambda^\alpha \tilde \Phi_\alpha \}_{D(\chi)}
\eeq  
where the Dirac bracket with respect to the second-class constraints 
$\{\chi_l\}$ is defined as 
\beq
\{ F, G \}_{D(\chi)} \equiv \{F, G\} - \{F, \chi_l \} \{\chi, \chi\}^{-1}_{l,l'} 
\{\chi_{l'}, G\} 
\eeq 
and $\t \Phi_\alpha$ and $\t \lambda_\alpha$ respectively denote the primary 
constraints 
which do not belong to the set $\{\chi_l\}$ and the corresponding 
Lagrange multipliers. In particular, when $\{\chi_l\}$ is the maximal set of 
second-class constraints in the system, $\t \Phi_\alpha$'s are all necessarily 
first class. In our case, $\{\t \Phi_\alpha\}= \{ \Omega, \Omega_a, \t\Omega_a, 
\pi^0 \}$. When any of the constraint depend on time, this formalism should be 
modified slightly. We should treat the time $\tau$ as additional canonical 
coordinate, and should introduce the conjugate momentum $\epsilon$\cite{GT}. The 
Poisson bracket should include the derivative with respect to these variables, 
and $H$  should be replaced by $H+\epsilon$.

\section{Counting of Degrees of Freedom}
The Hamiltonian formalism has an advantage that it exhibits the 
relevant dynamical degrees of freedom in a clear manner. It is quite well known 
that by an appropriate canonical transformation, the second class constraints 
can be identified with pairs of canonically conjugate pairs, $(P_\alpha, 
Q^\alpha)$, at least for bosonic theory\cite{GT,HT}. Again we see that the 
number of second-class constraints for bosonic theory must be even. For 
each pair of second-class constraints, there is a degree of freedom which is 
eliminated from the theory. The first-class constraints are written (after 
suitable canonical transformation) as a set of momenta $P_\beta$, and the 
coordinates $Q^\beta$ conjugate to these momenta are completely arbitrary. Since 
the time evolution of these variables are not determined from the dynamics, we 
interpret these variables as unphysical. One can in principle put conditions 
$Q^\beta=0$ by hand in order to make the first-class constraints into 
second-class ones. It is the gauge fixing. Therefore we see that for each 
first-class variables, one degree of freedom is removed. Therefore one has 
\cite{HT}:
\bea
\rm (\matrix{\hbox{Number of physical}\cr \hbox{degrees of 
freedom}})&=&(\matrix{\hbox{Total number of}\cr 
\hbox{coordinates}})-(\matrix{\hbox{Number of second-}\cr\hbox{-class 
constraints}})/2  \cr && - (\matrix{\hbox{Number of first-}\cr\hbox{-class 
constraints}})
\eea
  In our case, we see that all the components of the auxiliary field $k_{ij}$ 
are completely removed by the first-class constraints (\ref{f1}), (\ref{f2}) and 
the second-class constraints (\ref{dnt})-(\ref{yes}).  The first-class 
constraints (\ref{f5}),(\ref{f6}) removes $n$ degrees from $X^\mu$, and 
(\ref{f3}), (\ref{f4}) removes 2 degrees from $A_a$, so the number of physical 
degrees of freedom at each point of worldvolume is given by 
\beq
(D-n)+(n-2) = D-2
\eeq
where $D$ is the dimension of the target space. We observe that it is 
independent of the dimension of the brane. Were it an ordinary $(n-1)$-brane, we 
would have the first term $D-n$ only, since there is no gauge field. For a given 
target space dimensionality, $D-n$ indicates the number of transverse 
directions. For D-branes, as the dimensionality of the brane grows, the number 
of transverse directions decreases, however the components of gauge 
field increases, thus keeping the number of physical degrees constant.
\section{Gauge fixing}
 As discussed above, the first-class constraints can be made into the second 
class by putting additional constraints by hand, which is nothing but gauge 
fixing. For bosonic variables, the total number of additional constraints should 
be equal to that of the first-class constraints present if one wants to remove 
all the unphysical degrees of freedom, without imposing unnecessary restrictions 
on the physical ones (overfixing). There are several choices one can consider. 
The static gauge discussed in \cite{JSetal} has an advantage of fixing all the 
first-class constraints, but since one is identifying worldvolume ``time" $\tau$ 
with a target space-time $X^0$ instead of a light-cone coordinate, one 
eventually gets some complicated expression for the Hamiltonian which involves 
square root. On the other hand, in the covariant gauge discussed in \cite{Ka} 
the bosonic first-class constraints were not fixed at all, and 
the gauge-invariance condition was imposed on the physical state vector. Since 
the primary constraints appearing in $H^{(1)}$ still remains first-class, unless 
further gauge condition is imposed as constraints at the operator level, no 
meaningful expression for the Hamiltonian without Lagrange multiplier can be 
found.   
In this work, I impose similar conditions as the light-cone gauge for the 
string\cite{GS}, (and also for membranes\cite{BST}), along with the temporal 
gauge for the gauge field. I fail to make all the first-class constraints into 
second class, but we can take advantage of the fact that as long as we 
put conditions so as to make all the {\it primary constraints} second-class, we 
can still eliminate all the Lagrange multiplier from $H^{(1)}$\cite{GT,HT}. 
Also, leaving just right amount of first-class constraints allow us to simplify 
Dirac brackets, and the resulting Hamiltonian takes a rather simple form in the 
light-cone gauge, as we will see below. Because of these advantages, the 
light-cone gauge has been used not only for string, but also for 
membrane\cite{BST}. The conditions are  
\begin{mathletters}
\label{gfixing} 
\bea
G_1 &\equiv& X^+ - P^+_0 \tau \approx 0 \\
G_2 &\equiv& \cases{ P^+ - (P_0^+)^{n-1} \approx 0 & $(n \neq 1)$ \cr  P^+ - 
{P_0^+ \over N} \approx 0 & $(n=1)$ } \\
G &\equiv& N-\sqrt{\g}(P_0^+)^{2-n} \approx 0  \quad (n \neq 1) \\
G^a &\equiv& N^a+\tN^a \approx 0 \\
F &\equiv& A_0 \approx 0\\
\varphi_{a b} &\equiv& \partial_a X^\mu \partial_b X_\mu + F_{b a} - \gamma_{a
b} \approx 0 \quad (n=2) 
\eea 
\end{mathletters} 
where the light-cone index is defined by $V^{\pm} \equiv {V^0 \pm V^9 \over 
\sqrt{2}}$.
Requiring these constraints to be maintained in time does not lead to any 
new constraint and all the remaining primary inexpressible velocities 
are determined. This means that there is no ambiguity left in the time evolution 
of the system. As was discussed above, since the number of gauge-fixing 
conditions we put in does not match the number of first-class constraints, there 
are still $n-1$ local first-class constraints left however. These 
represent time-independent gauge transformation of the vector field and the 
volume-preserving diffeomorphisms, and are given by,
\begin{mathletters}
\label{resid}
\bea
&&\phi = \partial_a \pi^a \label{res} \\
&&\epsilon^{a_1 a_2 \cdots a_{n-1}}\nabla_{a_1}[ T_{a_2} +\sqrt{\g} 
\partial_{a_2} \Pi] \label{resi} 
\eea 
\end{mathletters}
 It is easily seen that $\phi$ is a first-class constraint by considering 
its Poisson bracket with other constraints. That (\ref{resi}) is first-class 
follows from the fact that the only nonvanishing Poisson bracket of 
$T_a+\sqrt{\g}\partial_a \Pi$ with  other constraint is given by 
\beq
\{ T_a + \sqrt{\g} \partial_a \Pi , P^+ - P_0^+ \} \approx P_0^+ 
\partial_a \delta (x-y) 
\eeq 
(\ref{resi}) form the divergence-free part of $T_a+\sqrt{\g} \partial_a \Pi$ 
and there are $n-2$ independent components among them.  For D-branes which are 
not simply connected, one expects that there are also global constraints which 
remain first class, (for membrane, see ref.\cite{BST}) but we will discuss only 
local properties for simplicity. Since we left some of the unphysical degrees of 
freedom, we have to make sure the physical quantities depend only on physical 
variables at the classical level. That is, for any physical quantity $F$, we 
should have 
\beq
\{ \Phi_\alpha, F  \}_D(\chi)=0
\eeq
where $\Phi_\alpha$ is any one of the remaining first-class 
constraints (\ref{resid}), and the Dirac bracket is evaluated using the 
second-class constraints. As will be shown in the next section, variables other 
than transverse components of $X^\mu$ and spatial components of $A_i$, 
respectively,  can be eliminated from the theory, and the Dirac bracket reduced 
to Poisson bracket, so one has
\beq
\{ \Phi_\alpha, F({\bf X, A})  \}=0
\eeq
\
where ${\bf X, A}$ represent transverse and spatial components of $X^\mu$ and 
$A_i$ respectively.  At the quantum level, the Poisson bracket above is simply 
replaced by the commutator. Also, at the quantum level, any physical state 
vector should satisfy the condition
\beq
\Phi_\alpha \vert phys > = 0,
\eeq  
That is, the physical wavefunction should depend only on physical variables.
 
\section{Light-cone Hamiltonian}
As was discussed in the last section, now we have a set of 
second-class constraints which are enough fix all the primary inexpressible 
velocities and the time evolution of the system is governed by $H$ in 
(\ref{ham}), 
which can be rewritten as
\beq
H =\int d^{n-1} \xi [ {N \over \sqrt{\g}}\varphi + {N^a + \tN^a \over 2} 
\varphi_a + A^0 \bar \phi ] \label{newham}
\eeq 
Since second class constraints can be set strongly equal to zero inside the 
Dirac bracket, we have $H=0$, and the whole dynamics are determined by 
the dependence of the constraints on the time $\tau$. One has
\beq
\dot \eta = \{ \eta, H+\epsilon  \}_{D(\chi)}=\{ \eta,  \epsilon  \}_{D(\chi)}
\eeq        
where all the Poisson brackets used in defining the Dirac bracket include 
the derivative with respect to $\tau$ and $\epsilon$. One can also make a 
time-dependent canonical transformation so that all the constraints become 
time-independent. The new Hamiltonian after this transformation is nonzero, 
which governs the time evolution of the system. Since we identify the time with 
one of the light-cone coordinates, this Hamiltonian is called the 
light-cone Hamiltonian. To be explicit, we make the canonical transformation 
$\bar X^+ = X^+ - P_0^+ \tau$ with other variables unchanged so as to make the 
time-dependent constraint $G_1=X^+-P_0^+ \tau$ into time-independent one $\bar 
G_1 = \bar X^+$. Then the corresponding generating functional $F$ has the form
\beq
F = X^\mu P_\mu -\tau P_0^+ P_+     
\eeq 
Denoting the new Hamiltonian after this transformation by $H^+$, we have 
\beq
H^+ = {\partial F \over \partial \tau} = -P_0^+ P_+
\eeq 
and  all the second class constraints can be rewritten in equivalent form 
which are paired as follows:
\bea
&&\cases{N-\sqrt{\g}(P_0^+)^{2-n} & \cr \Pi& } \qquad \cases{
A_0 &\cr \pi^0} \cr
&&\cases{N^a+{N \over \sqrt{\g}}\pi^a &\cr  \Pi^a & } \quad
\cases{\tN^a-{N \over \sqrt{\g}}\pi^a &\cr  \tP^a & } \cr
&&\cases{\g_{ab}-\partial_a X^\mu \partial_b X_\mu - F_{ba}&\cr \Pi^{ab}& } \cr
&&\cases{\bar X^+  &\cr P_+ - {1 \over  2 P_0^+}[{\bf P}^2+\g(\g^{cd}  +{n_-^c 
n_-^d \over 4N^2})
\partial_c X \cdot \partial_d X  - \g  (n-2)  + \g \g^{cd} F_{cd} ]&} \cr
&&\cases{X^- -{1 \over P_0^+ \nabla^2} [ \nabla^a({\bf P} \cdot \partial_a {\bf 
 X})- 
\nabla^a\{{\sqrt{\g}\over 2N} F_{ab} n_-^b \}]&\cr  P_- - (P_0^+)^{n-1} &\cr}  
\eea 
where $\bf X,P$ indicate the vectors consisting of transverse components 
$X^l,P_m \ (l,m=1 \cdots 8)$ only. When $n=1$, $N-\sqrt{\g}$ in the first pair 
is replaced by $N-{P \over P_0^+}$ and the last pair is removed.
As is discussed in Appendix, these constraints are of the special form and the 
 canonical variables $(N,\Pi), (N^a,\Pi_a), (\tN^a,\tP_a),(\g_{ab}, 
\Pi^{ab}),(X^+, P_+), (X^-, P_-), (A_0, \pi^0)$ can be solved in terms of the 
other variables $(X^l, P_m)\ (l,m=1, \cdots 8), (A_a, \pi^a)$ using the 
constraints above and the Dirac brackets between these remaining variables 
reduces to the Poisson brackets.  Therefore,  we finally have the light-cone 
Hamiltonian
\beq 
H^+ = {1 \over 2} [ {\bf P} ^2 + \det(\partial_a {\bf X}  \cdot \partial_b 
{\bf X} + F_{ba}) + \pi^c \pi^d \partial_c {\bf X}  \partial_d {\bf X}]
\eeq 
and the equation of motion 
\beq
\dot \eta^* = \{ \eta^*, H^+ \}
\eeq 
where $\{\eta^*\}$ are the remaining canonical variables $X^l, P_m, A_a,\pi^b$.
 
\section{Summary} 
 In this paper we considered the Hamiltonian formalism for the D-brane action 
 which is written in terms of auxiliary fields. We introduced the 
generalized shift and lapse functions, which made it very natural to go to the 
light-cone gauge. constraint analysis was done in detail for the case of bosonic 
D-brane. We then derived the light-cone Hamiltonian which is quadratic in 
field momenta. Although not done in this paper, one might also apply similar 
analysis for supersymmetric or non-Abelian cases. After this paper was completed 
and submitted, ref.\cite{MMM} appeared in the preprint archive, where the 
light-cone formulation for bosonic D2-brane is also discussed.     

\acknowledgements
The author would like to thank Profs. Yongmin Cho and John H. Schwarz for 
discussions. This work is supported in part by Asia Pacific 
Center for Theoretical Physics  and in part by Korea Research Foundation through 
Research Institute for Natural Sciences, Hanyang University.

\appendix
\section*{Appendix: Constraints of special form and simplification of 
Dirac brackets} 
Let a complete set of second class constraints be given by $\chi$. Let us solve 
the equations $\chi=0$ with 
respect to some variables $\tilde \eta$, which are a set of pairs of 
canonical variables. We denote the rest of the pairs by $\bar \eta$, and the 
canonical variables are given by $\eta=(\tilde \eta, \bar \eta)$.  Since we may 
use any equivalent set of constraints for evaluating Dirac 
brackets\cite{GT,HT,D}, we assume without loss of generality that $\chi$  takes 
the form $\chi=\tilde \eta - f(\bar \eta) \label{cons}$,  
that is, $\chi$ can be used to eliminate variables $\tilde \eta$ 
 in terms of $\bar \eta$.  
  Since the constraints can be set to zero inside the Dirac 
bracket\cite{GT,HT,D}, we have
\beq
\dot {\bar \eta} =  \{\bar \eta, \bar H\}  -  \{\bar \eta, 
\chi_i\} (\{ \chi, \chi\} ^{-1}_{ij}\{\chi_j, 
\bar H\}  \label{eom} 
\eeq 
where $\bar H \equiv H(\tilde \eta(\bar \eta), \bar \eta)$, i.e. the Hamiltonian 
expressed in terms of $\bar \eta$ only.
 The discussion so far is quite general. However, the equation of motion 
is considerably simplified when $\chi$ are of the forms so that the second 
term  on the right-hand side of (\ref{eom}) vanishes. Such constraints $\chi$ 
will be called {\it constraints of special form}\cite{GT}. In particular,   
we consider the case where the constraints are paired in the form
\bea
\tilde q^n-{\rm constant}, &&\; \tilde p_n - f^{2n}(\bar \eta), \quad {\rm or} 
\nonumber \\
\tilde q^n - f^{1n}(\bar \eta), &&\; \tilde p_n - {\rm constant} . 
\eea 
Then the matrix $\{\chi,\chi\}$ are written in a block diagonal 
 form:
\beq
\{\chi,\chi\}=\pmatrix{0&I&0&0\cr
-I&\{f^{2n},f^{2n'}\}_{D(\psi)}&\{f^{2n},f^{1n}\}_{D(\psi)}&0\cr
0&\{f^{1n},f^{2n}\}_{D(\psi)}&\{f^{1n},f^{1n1}\}_{D(\psi)}&I\cr
0&0&-I&0  }
\eeq 
and the inverse matrix is
\beq
(\{\chi,\chi\})^{-1}=\pmatrix{\{f^{2n},f^{2n'}\}_{D(\psi)}&-I&0&
 -\{f^{2n},f^{1n}\}_{D(\psi)}\cr
I&0&0&0\cr
0&0&0&-I\cr
\{f^{1n},f^{2n}\}_{D(\psi)}&0&I&\{f^{1n},f^{1n1}\}_{D(\psi)} }
\eeq 
It is easy to see that indeed in this case the second term of the right-hand 
side of (\ref{eom}) vanishes and the Dirac bracket simply reduces to the 
 Poisson bracket. 
\end{document}